# High-resolution deuterium metabolic imaging of the human brain at 9.4 T using bSSFP spectral-spatial acquisitions

Praveen Iyyappan Valsala[1,2*], Rolf Pohmann[1], Rahel Heule[1,2,3], Georgiy A. Solomakha[1], Nikolai I. Avdievich[1], Jörn Engelmann[1], Laura Kuebler[4,5,6], André F. Martins[4,5,6], Klaus Scheffler[1,2]

[1]High Field Magnetic Resonance, Max-Planck Institute for Biological Cybernetics, Tübingen, Germany

[2]Department of Biomedical Magnetic Resonance, Eberhard Karls University Tübingen, Tübingen, Germany

[3]Center for MR Research, University Children's Hospital, Zurich, Switzerland

[4]Werner Siemens Imaging Center, Department for Preclinical Imaging and Radiopharmacy, Eberhard Karls University Tübingen, Tübingen, Germany

[5]Cluster of Excellence iFIT (EXC 2180) «Image-Guided and Functionally Instructed Tumor Therapies», Eberhard Karls University Tübingen, Tübingen, Germany

[6]German Cancer Consortium (DKTK), partner site Tübingen, German Cancer Research Center (DKFZ), Im Neuenheimer Feld 280, Heidelberg 69120 (Germany)

*Praveen Iyyappan Valsala (praveen.valsala@tuebingen.mpg.de) is the corresponding author

# Abstract

## Purpose
To improve the sensitivity and robustness against $B_0$ inhomogenities of deuterium metabolic imaging with balanced steady-state methods at 9.4 T.

## Methods
We investigated two variants of bSSFP acquisitions, namely uniform-weighted multi echo and acquisition-weighted CSI to improve the SNR of deuterium metabolic imaging (DMI) in the brain with oral [6,6'-$^2$H$_2$]-glucose intake. Phase-cycling was introduced to make bSSFP acquisitions less sensitive to $B_0$ inhomogeneity. Two SNR optimal methods for obtaining metabolite amplitudes from the phase-cycled data were proposed. The SNR performance of the two bSSFP variants was compared with a standard gradient-spoiled CSI acquisition and subsequent IDEAL processing. In addition, *in vivo* $T_1$ and $T_2$ of water, glucose and Glx (glutamate+glutamine) were estimated from non-localized inversion recovery and spin-echo measurements.

## Results
High-resolution whole-brain dynamic quantitative DMI maps were successfully obtained for all three acquisitions. Phase-cycling improved the quality of bSSFP metabolite estimation and provided additional spectral encoding. The SNR improvement was only observed for the CSI variant of bSSFP acquisitions with an average increase of 18% and 27% for glucose and Glx, respectively, compared to the vendor's CSI. ME-bSSFP acquisition achieved higher resolutions than acquisition-weighted CSI and exhibited several qualitative improvements.

## Conclusion
We demonstrated the feasibility of using CSI-bSSFP and ME-bSSFP acquisitions for off-resonance insensitive high-resolution [6,6'-$^2$H$_2$]-glucose DMI studies in the healthy human brain at 9.4T. Balanced SSFP acquisitions have potential to improve the sensitivity of DMI despite the SNR loss of phase-cycling and other human scanner constraints.

# Introduction

Deuterium metabolic imaging (DMI) has recently gained traction in the field of metabolic research for the study of normal and diseased metabolic pathways in pathologies such as cancer, diabetes, and neurodegenerative diseases[1,2]. Due to the low natural abundance of deuterium in the human body, metabolic studies are performed with deuterated tracers. Most studies use [6,6'-$^2$H$_2$]-labeled glucose, which provides comprehensive information on glucose uptake, glycolysis and the tricarboxylic acid cycle (TCA) cycle. In contrast, the routinely used fluorodeoxyglucose (FDG) positron emission tomography (PET)[3] in clinical practice only provides information on glucose uptake. Furthermore, the negligible natural abundance and the few resonances of the exogenous tracer and its downstream metabolites enables straightforward acquisition and dynamic metabolite concentration estimation without many confounders[4]. This offers an advantage over other experimental MR metabolic imaging methods, including $^1$H MRSI[5], $^{13}$C MRSI[6] and glycoCEST[7]. The non-radioactive deuterium tracers are also suitable for longitudinal studies as they can be safely taken orally.

Despite its promises, the low sensitivity of DMI resulting from the low gyromagnetic ratio and low concentration of deuterated metabolites limits its practical applicability. DMI studies are predominantly conducted at high field strengths because the signal-to-noise-ratio (SNR) increases supra-linearly with the static magnetic field[8] ($SNR \propto B_0^{1.7}$). Glucose-DMI investigations in the human brain have reported spatio-temporal resolutions ranging from 3 mL in 10 min at 9.4 T to 14 mL in 39 min at 3 T[2,9-11]. The acquisition weighting used in all these studies to improve SNR further degrades the previously mentioned nominal resolution by several orders of magnitude. Therefore, improving the sensitivity of DMI is highly desirable. Recently, balanced steady state free precession (bSSFP) acquisition[12] rather than RF-spoiled gradient-echo (FLASH) acquisition has been proposed in a 15.2 T pre-clinical DMI study [13,14]. The rationale behind was the much higher T2/T1 ratio of deuterated metabolites compared to protons, even at high field strengths, yielding considerable signal enhancement when using bSSFP acquisitions[15].

In this clinical translational study, the feasibility of using bSSFP acquisitions to improve [6,6'-$^2$H$_2$]-glucose DMI in the healthy human brain at 9.4 T is investigated. To this end, bSSFP sequences employing two popular 3D encoding schemes, namely chemical shift imaging (CSI) and spectroscopic multi echo, were compared. For both bSSFP acquisition types, phase cycling was incorporated to enhance robustness against the substantial B$_0$ inhomogeneity in the human head. Two spectral fitting methods were developed for the phase-cycled bSSFP data to obtain reliable and SNR optimal metabolite maps. The SNR performance and spectral-spatial encoding of these two bSSFP acquisition schemes were compared to the vendor's standard implementation consisting of a gradient spoiled 3D CSI sequence without RF-spoiling (FISP/lowest-order SSFP-FID)[16]. The DMI protocols were designed to maximize SNR based on the simulation of the acquired signal utilizing the measured *in vivo* T1, T2 and T2* relaxation times. The spectral separation of the designed protocols was verified using a phantom containing all four (water, glucose, glutamic acid and lactate) deuterated compounds detected in *in vivo* [6,6'-$^2$H$_2$]-glucose DMI[4].

# Methods

## Simulation

The SNR improvements for different metabolites were simulated with the analytical signal equations of FLASH[17], FISP[18] and bSSFP[19] acquisition methods. The measured *in vivo* relaxation times ($T_1, T_2, T_2^*$) of the metabolites and system constraints (energy deposition limit and $B_0$ inhomogeneity) were included to estimate the signal dependence as realistic as possible. The effects of $T_2^*$ decay during signal readout, ADC (analog-to-digital converter) duty cycle and phase-cycling during bSSFP acquisition was modelled into the signal equation. In addition, the noise amplification during spectral separation for a given echo spacing/readout duration was also optimized by maximizing number of signal averages[20].

## Measurement setup

All measurements were performed on a 9.4 T Siemens Magnetom whole-body human scanner. A double-tuned phased array radiofrequency (RF) coil with 8 TxRx/2Rx deuterium channels and 10 TxRx proton channels [9,21] was used. The new software baseline (VE12U) of the scanner natively supports deuterium imaging with an additional license.

## Phantom

To investigate the SNR improvements and spectral resolution of different acquisition methods, we constructed a 18 cm diameter spherical phantom with nine 20 mL vials containing different metabolite mixtures (deuterated water, [6,6'-$^2$H$_2$]-glucose, [4,4'-$^2$H$_2$]-glutamic acid and [3,3,3-$^2$H$_3$]-lactate). The vials were distributed along the outer periphery to minimize bias due to transmit and receive field inhomogeneity. For gelation in both phantoms and vials, we used 3.1 % w/w hydroxyethylcellullose which facilitates air bubble removal even after hydrogel formation. A detailed description of the phantom construction is provided in the shared data repository.

## Human Subjects

Five healthy subjects (age = 32±5 years, 2 males) took part in the study and were asked to fast overnight and ingest [6,6'−$^2$H$_2$] glucose solution (0.75g / kg body weight) prior to the measurements. All *in vivo* experiments were approved by our local ethics committee and were only performed after obtaining written consent from the study participants. The final optimized protocols were tested on three subjects, while the data from two additional subjects was used for non-localized relaxometry and protocol optimization. The measurements were conducted over a period of approximately two hours, during which all three DMI protocols described below were interleaved at least twice. In addition, non-localized relaxometry and structural scans were performed.

## DMI protocols

All *in vivo* deuterium imaging protocols were designed for 10 minutes of acquisition time irrespective of the resolution and other acquisition parameters to facilitate comparison. The important acquisition parameters of all spectral-spatial $^2$H imaging sequences are summarized in Table 1. The CSI-FISP sequence was optimized for SNR with an optimal readout length of 1.26 times $T_2^*$, assuming a $T_2^*$ of 22 ms[22]. All protocols used non-selective rectangular RF pulses to improve maximize ADC duty cycle and minimize specific absorption rate (SAR) restrictions. The readout length of 14.9 ms for CSI-bSSFP and 5 echoes with an echo spacing of 3.4 ms for ME-bSSFP was sufficient to achieve spectral separation of all four deuterated metabolites at 9.4 T. Mono-polar readout gradients were used in ME-bSSFP due to the low bandwidth readout to ensure unidirectional pixel shifts. To increase the number of signal averages, both CSI protocols employed Hamming acquisition weighting while ME-bSSFP employed elliptical scanning. The resolution loss due to Hamming acquisition weighting and the elliptical k-space sampling was estimated from the full width of point spread function (PSF) at 64% peak height. In addition to the standard protocols, an adapted protocol without phase-cycling was used in a phantom measurement to demonstrate off-resonance insensitivity and additional spectral encoding provided by phase cycling.

| protocol | FOV [cm] | Matrix size | TR [ms] | Repetitions (averages/ phase cycles) | RF pulse FA [deg] | RF pulse Dur [ms] | ADC duty cycle [%] | Nominal voxel volume [ml] | PSF voxel volume [ml] |
|---|---|---|---|---|---|---|---|---|---|
| CSI-FISP | 20x20x20 | 24x24x24 | 36 | 12/n.a | 41 | 0.5 | 79 | 0.58 | 2.85 |
| CSI-bSSFP | 20x20x200 | 24x24x24 | 19 | 6/4 | 50 | 1.4 | 78 | 0.58 | 2.98 |
| ME-bSSFP | 40x20x30 | 32x16x24 | 19 | 5/18 | 50 | 1.4 | 71 | 1.95 | 2.18 |

**Table 1:** Main sequence parameters of all three DMI protocols used in this study. **FOV**, field of view; **TR**, repetition time; **FA**, flip angle; **Dur**, RF pulse duration; **PSF**, point spread function; **ADC**, analog to digital

### Image reconstruction

The acquired data was noise-prewhitened, averaged, zero-padded to double the encoding matrix size and reconstructed using a 3D fast Fourier transform (FFT) along the three physical dimensions to obtain the reconstructed volume at different time points and phase cycles. After reconstruction, adaptive coil combination[23] was used for both CSI and multi-echo data to combine multi-channel images. The coil weights were calculated by averaging the image volume across the echo-time and phase-cycling dimension.

### Spectral fitting

The metabolite amplitudes from the CSI-FISP datasets were estimated by simultaneously fitting the phase evolution of the metabolites along the readout and $B_0$ off-resonance dimension using the standard IDEAL algorithm[24]. In the case of phase-cycled bSSFP data, we propose two methods to handle the additional amplitude and phase modulations introduced by the phase-cycling: IDEAL-modes fit and linear fit. The key steps involved in those two proposed methods are schematically illustrated in Figure 1. The IDEAL-modes fit employed a relatively simple phase evolution signal model similar to the standard IDEAL algorithm in the SSFP configuration space[12] followed by a data-driven approach for optimally combining metabolite amplitudes across SSFP modes to maximize SNR. In the linear fit, phase-cycling related amplitude and phase modulations in the data were explicitly formulated in the signal model to provide additional information for spectral separation. The $B_0$ off-resonance map used for both methods was estimated from the lowest configuration order ($F_0$ mode) using the IDEAL algorithm[24].

In the IDEAL-modes fit, SSFP configuration modes were estimated via a discrete Fourier transform (DFT) of the acquired phase cycles to remove the off-resonance-dependent spatial amplitude variations (i.e. bandings). As a result of the finite sampling of the bSSFP frequency response, the highest SSFP configuration order C that can be estimated with K phase cycles is given by C=K/2-1. Extending Equation [10] from Nyugen and Bieri[25] for M metabolites, the SSFP mode amplitude $F(c, t_n)$ of an image pixel at a discrete time point $t_n$ (n=1,...,N) and configuration order c (c = -C,...,0,...,C), by ignoring T$_2$ relaxation effects is given by

$$F(c, t_n) = \sum_{j=1}^{M} P_{j,c} e^{i\varphi_j(t_n)} e^{ic\Phi_j}$$

where $P_{j,c}$ is the j[th] metabolite amplitude for SSFP configuration order c. The phase terms $\varphi_j(t_n) := (t_n/TR)\Phi_j$ and $\Phi_j := 2\pi(\Delta f_j + \Delta f_0)TR$ refer to the phase accumulated by the j[th] metabolite at a discrete echo time $t_n$ and the phase accumulated during each repetition time (TR), respectively. The off-resonance terms $\Delta f_j$ and $\Delta f_0$ refer to the chemical shift of the j[th] metabolite and the B$_0$-related off-resonance in Hz at the spatial position, respectively. The complex weights for optimally combining metabolite amplitudes $P_{j,c}$ of all the modes was obtained for individual metabolites by estimating the principal eigenvector with the largest eigenvalue. Because of the flexibility of this method and slightly better SNR performance than the linear fit method, it was used throughout this work unless specified otherwise.

In the linear fit, the acquired phase-cycled bSSFP data were fitted to bSSFP frequency responses, simulated with a more complete signal model[19] than the simple phase evolution model used in the IDEAL-modes approach, using a linear least squares algorithm to estimate the metabolite amplitudes directly. The bSSFP signal $S_j$ of the j[th] metabolite with chemical shift $\Delta f_j$ at discrete time point $t_n$ (n=1,...,N) was calculated by adding phase evolution and relaxation terms during free induction decay to the transverse magnetization after excitation as described by Ganter in equations [42-44][26]. In addition to the chemical shift $\Delta f_j$ and time point $t_n$, the signal depends on off-resonance $\Delta f_0$, T$_1$ and T$_2$ relaxation times, as well as acquisition parameters (flip angle $\alpha$, RF phase increment $\psi$, repetition time $TR$):

$$S_j(t_n, \psi) = \frac{(1 - E_{1,j})\sin\alpha \left(1 - E_{2,j}e^{-i\theta_j}\right)\left(e^{-\frac{t_n}{T_{2,j}}}e^{i\varphi_j(t_n)}\right)}{(1 - E_{1,j}\cos\alpha)(1 - E_{2,j}\cos\theta_j) - (E_{1,j} - \cos\alpha)(E_{2,j} - \cos\theta_j)E_{2,j}}$$

with $E_{1,j} := \exp\left(-\frac{TR}{T_{1,j}}\right)$ and $E_{2,j} := \exp\left(-\frac{TR}{T_{2,j}}\right)$ referring to the relaxation terms of the j[th] metabolite, and $\theta_j := \Phi_j - \psi$ denoting the phase difference between the off-resonance-related phase term Φ_j:=2π(Δf_j+Δf_0 )TR accumulated during each TR interval and the RF phase increment ψ, as well as φ_j (t_n ):=(t_n/TR) Φ_j  denoting the phase accumulated by the transverse magnetization at a discrete time t_n.

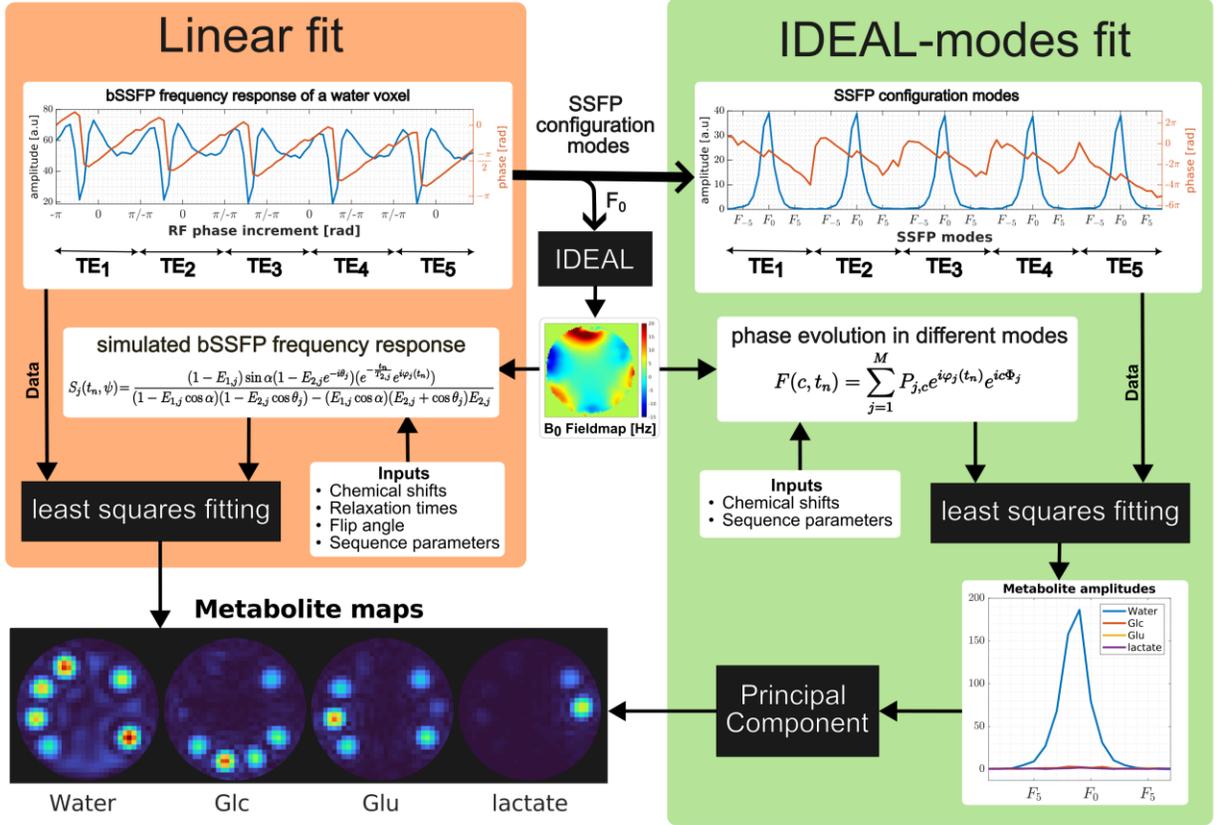

**Figure 1:** Overview of the two proposed data processing methods for phase-cycled bSSFP data. An exemplary bSSFP frequency response of a water voxel from a ME-bSSFP phantom experiment for all echo times is fitted to the simulated bSSFP frequency response using the linear fit method. For the IDEAL-modes method, only the phase evolution is fitted to the individual SSFP configuration modes and the metabolite amplitudes across different SSFP modes are optimally combined to obtain final metabolite maps. Both methods rely on the $B_0$ field map estimated from complex-averaged phase-cycled bSSFP data (lowest-order SSFP-FID/$F_0$ mode) at multiple echo times obtained by the IDEAL algorithm.

### SNR calculation

For accurate SNR estimation, reconstruction was performed in SNR units[27] by using appropriate scaling factors during averaging, Fourier transform and coil combination after noise decorrelation. Additional correction for the anti-aliasing filter was not necessary because readout oversampling was removed from both noise and image data. The coil normalization[28] weights estimated during the coil combination step to correct for inhomogeneous receive sensitivity was not applied during SNR analysis. The Euclidean norm of the complex linear combination coefficients of the individual metabolites during spectral separation was used to scale the resulting metabolite maps to SNR units respectively. The final magnitude metabolites maps were scaled by a factor $\sqrt{2}$ to maintain the unit standard deviation of the noise floor. Additionally, the SNR maps were normalized with the estimated PSF voxel volume for fair comparison.

## Metabolite quantification

*In vivo* quantitative glucose and Glx (glutamate+glutamine) metabolite maps were obtained using the assumed homogenous water reference of 10.12 mM throughout the brain as shown by Peters et al [13]. The theoretical signal models referenced in the section Simulation were used for calibrating the metabolite amplitude ratios. Additionally, 33% label loss in Glx transformation during oxidative metabolism was considered[29]. In case of multi-echo acquisitions, the voxel shifts along frequency encoding direction due to chemical shifts during the low-bandwidth readout was estimated and corrected prior to metabolite quantification. The calculated voxel shifts were 1.8 mm, 4.8 mm and 6.8 mm for Glucose, Glx and lipid/lactate, respectively, if the system frequency is set to the water resonance.

## Non-localized spectroscopy

Non-localized inversion-recovery and spin-echo sequences written in the Pulseq framework[30] were performed to measure $T_1$ and $T_2$ relaxation times both *in vivo* and *in vitro*. For $T_1$ relaxometry, longitudinal relaxation was sampled at 15 inversion times ranging from 10 to 1500 ms with a repetition time of 2.5 s. A 5.2 ms non-selective adiabatic frequency offset corrected inversion (FOCI) pulse[31] was used as the inversion pulse. In case of $T_2$ relaxometry, non-selective block pulses with durations of 1 ms and 2 ms were used for excitation and refocusing, respectively, to achieve short echo times. The echo times were distributed between 10 and 700 ms with a repetition time of 2.5 s. Both $T_1$ and $T_2$ measurements took 10 minutes each with 16 averages for each inversion/echo time. The measurements were performed approximately 90 minutes after the labelled glucose intake in case of the *in vivo* scans.

The relaxometry data was noise decorrelated and coil-combined using the weighted singular value decomposition (WSVD) method[32]. The complex FID was fitted using a non-linear least squares algorithm called the AMARES algorithm[33,34] to estimate the peak amplitudes. The peak amplitudes from different inversion times (TI) and echo times (TE) were fitted to the standard exponential models $S(TI) = a(1 - 2e^{\left(\frac{-TI}{T_1}\right)}) + c$ and $S(TE) = ae^{\left(\frac{-TE}{T_2}\right)} + c$ to estimate the $T_1$ and $T_2$ relaxation times respectively. A bi-exponential fit was only required for the *in vivo* water $T_2$ relaxation fit to account for ventricular water[2]. Metabolite $T_2^*$ relaxation times and chemical shifts were obtained from the linewidth and off-resonance shift of the metabolite peaks in the CSI-FISP data estimated by the AMARES algorithm[33,34]. In addition to relaxometry, non-localized spectra were acquired at 10 different pulse powers to determine the actual reference power.

# Results

## Non-localized spectroscopy

The $T_1$ and $T_2$ relaxation times estimated from the non-localized inversion-recovery and spin-echo scans are shown in Table 2 for *in vivo* (n=5) and phantom experiments. The water resonance of the inversion-recovery data shows slight deviation from the Lorentzian line shape possibly due to the strong $B_0$ inhomogeneity at 9.4 T (see Figure S1). The spectral-fitting quality of inversion-recovery and spin-echo data acquired in a representative subject can be found in Supporting Information Figures S1 and S2, respectively. The assumed fraction of the ventricular water compartment with longer $T_2$ was estimated to be 11 ± 2% by bi-exponential $T_2$ relaxation fitting. The chemical shifts of the $^2H$ resonances obtained for the phantom data show an upfield shift of approximately 0.2 ppm with respect to water compared to *in vivo*.

|   | Resonances | $T_1$ [ms] | $T_2$ [ms] | $T_2^*$ [ms] | chemical shift [ppm] |
|---|---|---|---|---|---|
| *In vivo* | Water | 370 ± 27 | 33 ± 3 | 21 ± 2 | 4.70 |
|   | Water 2(11 ± 2%) | n.a | 311 ± 64 | n.a | n.a |
|   | Glucose | 78 ± 8 | 37 ± 3 | 14 ± 2 | 3.80 |
|   | Glx | 162 ± 7 | 97 ± 11 | 23 ± 1 | 2.35 |
|   | Lactate/lipids | 137 ± 22 | 93 ± 88 | 24 ± 5 | 1.37 |
| Phantom | Water | 457 ± 24 | 269 ± 4 | 75 ± 39 | 4.70 |
|   | Glucose | 60 ± 12 | 57 ± 4 | 22 ± 3 | 3.62 |
|   | Glutamic acid | 165 ± 18 | 112 ± 6 | 53 ± 14 | 2.16 |
|   | Lactate | 245 ± 12 | 222 ± 12 | 76 ± 29 | 1.18 |

**Table 2:** Relaxation times and chemical shifts of deuterium resonances for *in vivo* (n=5) and phantom data. $T_1$ and $T_2$ relaxation times are derived from non-localized inversion recovery and spin-echo acquisitions. Water 2 corresponds to the ventricular water compartment modelled during $T_2$ fitting. Median $T_2^*$ and chemical shifts are estimated from the CSI-FISP data using the AMARES algorithm.

## Simulation

Figure 2 shows the relative signal efficiency ($\frac{signal\ amplitude}{\sqrt{TR}}$) increase of the phase-cycled bSSFP acquisition compared with the FISP and FLASH acquisition for *in vivo* water, glucose and Glx relaxation times. According to the simulations, the expected SNR increase of bSSFP is 7%, 45% and 69% for water, glucose and Glx, respectively, as compared to the FISP acquisition. This comes in addition to the 5-10% SNR boost for all metabolites when using FISP acquisition instead of RF-spoiled FLASH acquisition. These predictions are based on the bSSFP-CSI protocol with 4 phase cycles and a duty cycle of 78%. A similar simulation for all four metabolites including lactate with the relaxation times measured in the phantom is included in Supporting Information Figure S3. For the relatively large $T_2/T_1$ ratios and long $T_2$ relaxation times of water and lactate in the phantom, an SNR increase of over 2 times is predicted while the improvement for glucose and glutamic acid is similar to the *in vivo* levels.

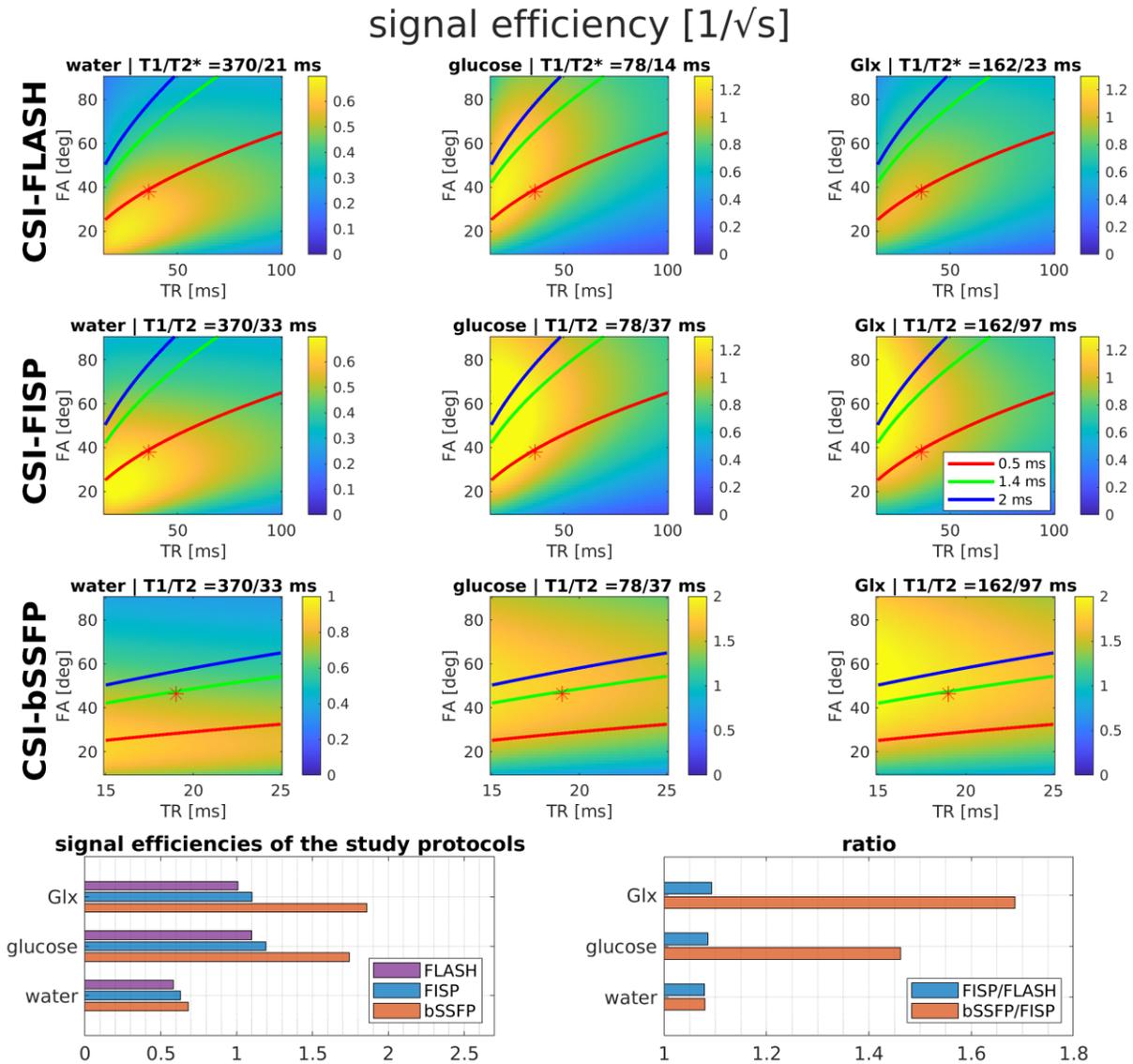

**Figure 2:** *In vivo* signal efficiency (signal amplitude/$\sqrt{TR}$) of deuterated metabolites with respect to TR and flip angle for FLASH, FISP and bSSFP acquisitions. The measured *in vivo* relaxation times used for the simulation are shown in the title of the respective subplots. The SAR limits with respect to TR for three pulse duration are overlaid to show the available parameter space for SNR optimization. The asterisks in the plots indicate the parameter combination (flip angle, TR) of the protocols used in this study. The signal efficiency and their ratios of the three protocols for water, glucose and Glx are depicted in the bottom panels.

## Phantom studies

In Figure 3, additional spectral encoding information in the phase cycle dimension is demonstrated by obtaining good spectral separation of all four resonances even with fewer than four echoes when phase cycling is performed. The condition number of the system matrix is lower and shows less dispersion due to $B_0$ inhomogeneity with more phase cycles. This minimizes additional noise amplification during spectral unmixing. The artefacts due to strong B$_0$ inhomogeneity in the water and glucose maps are reduced with more phase cycles.

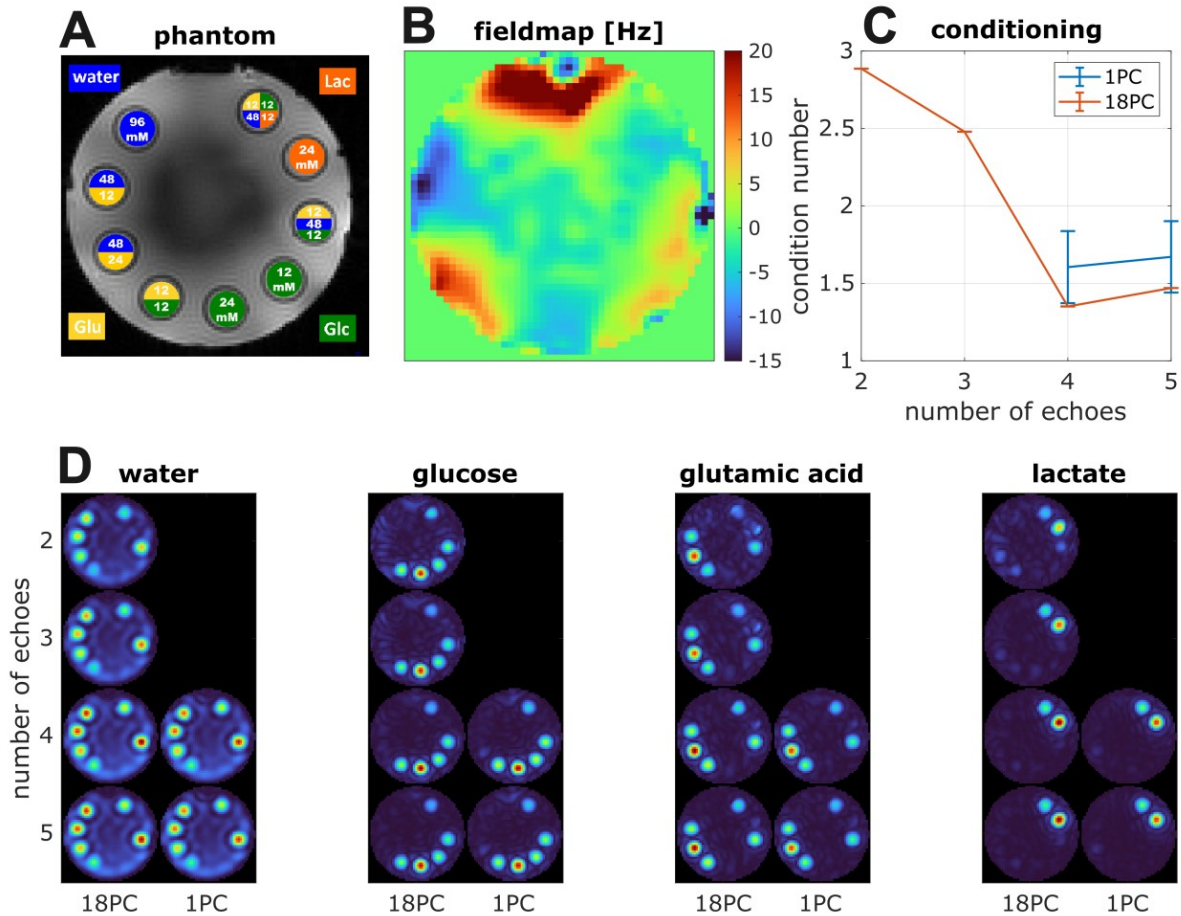

**Figure 3:** A) The constructed phantom with various concentrations of four deuterated compounds (blue: water, green: Glucose ,yellow: Glutamic acid, orange : lactate). B) shows the $^2$H B$_0$ map estimated using the IDEAL algorithm. C) shows better conditioning of the model for matrix inversion with higher number of phase cycles and echoes. The error bars indicate less dispersion due to B$_0$ off-resonance with 18 phase cycles. D) Metabolite maps from the ME-bSSFP protocols with 18 and 1 (RF phase increment of 180 deg) phase-cylce/s and different numbers of echoes (echoes were retrospectively removed). The four metabolites cannot be separated with less than four echoes in the data without phase-cycling.

The results of the SNR improvement investigation in the phantom for all metabolites using the *in vivo* protocols are shown in Figure 4. The resulting SNR metabolite maps were scaled with the relative PSF voxel volume from Table 1 for easy comparison. All three protocols exhibit excellent spectral separation of all four metabolites. An improvement in SNR for all four metabolites water, glucose, glutamic acid and lactate by 45%, 6%, 14% and 45%, respectively, is observed for the CSI-bSSFP acquisition as compared to the standard CSI-FISP acquisition. However, only a marginal SNR increase for water and lactate can be seen in the ME-bSSFP maps relative to the CSI-FISP maps. The other metabolites glucose and glutamic acid show a reduction in SNR by 26%, and 40%, respectively, for ME-bSSFP relative to CSI-FISP.

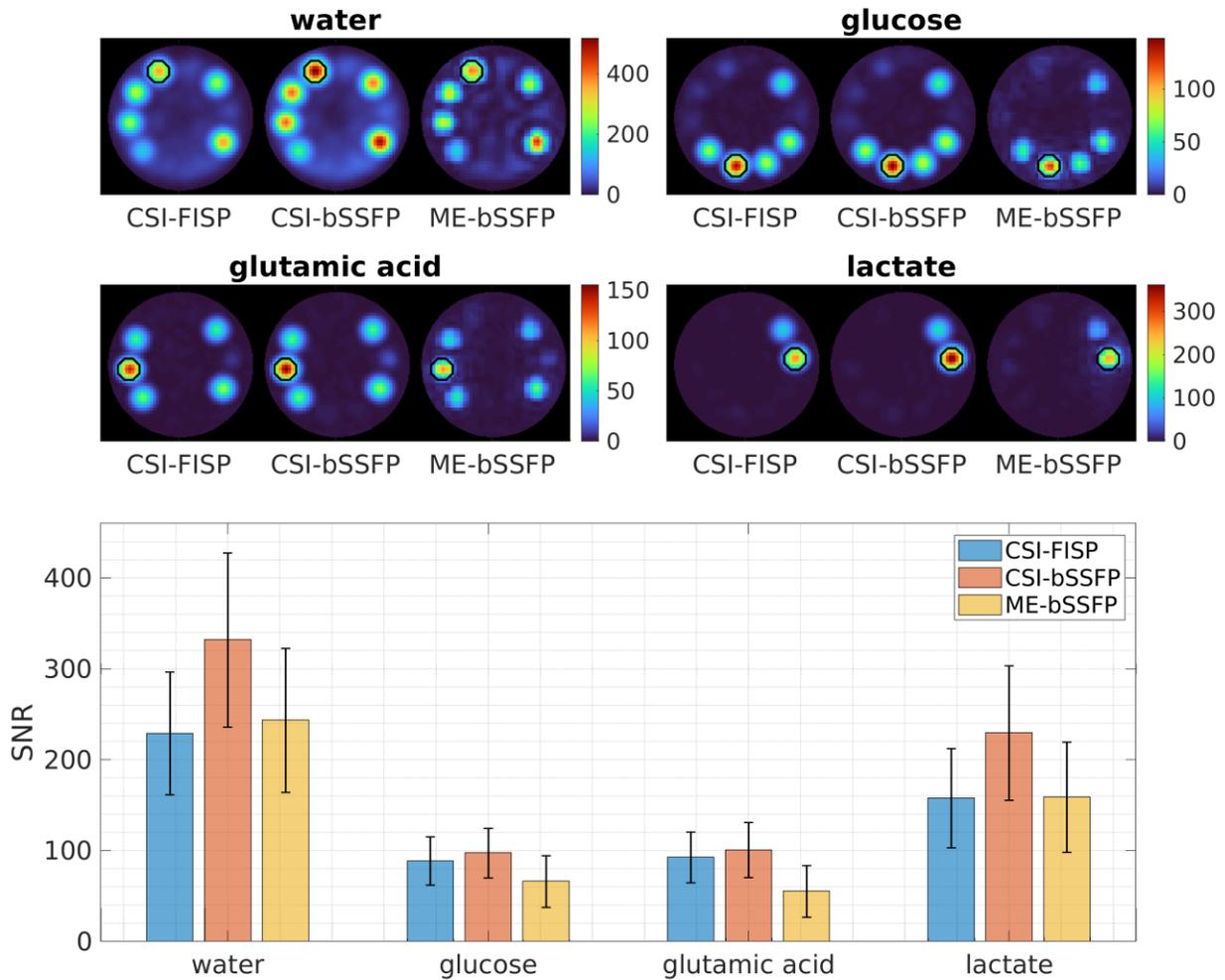

**Figure 4:** *In vitro* SNR analysis. The metabolite maps in SNR units obtained using all three investigated protocols are shown for the phantom study. The SNR maps are normalized with the PSF voxel volume of the acquisitions for fair comparison. The bar plots show the mean and standard deviation of the SNR within regions-of-interest (ROIs). The vials with the largest concentrations are chosen as ROIs. ROIs are indicated in the metabolite maps with black overlaid circles.

### *In vivo* DMI studies

Figure 5 shows the overview of *in vivo* spectral separation and SNR performance of all three acquisition protocols. A representative axial slice of the SNR maps and the whole-brain SNR distribution of all metabolites are presented for all subjects. The SNR maps from the interleaved DMI scans were averaged according to the acquisition type to minimize the impact of the temporal evolution of metabolites. The SNR maps from CSI-bSSFP and ME-bSSFP were scaled by a factor of 0.96 and 1.31, respectively, to account for the PSF voxel volume for quick comparison. An average SNR increase of 9-29% for Glucose and 27-33% for Glx is observed for the CSI-bSSFP acquisition in comparison to CSI-FISP. The water SNR shows only marginal differences between the two acquisitions. ME-bSSFP, on the other hand, exhibits a decrease in SNR by 20-40% for water and glucose relative to CSI-FISP, whereas the SNR for Glx is only marginally reduced. In subject 2, artefacts in the metabolite maps due to incorrect $B_0$ prediction by the IDEAL algorithm in the frontal region are evident.

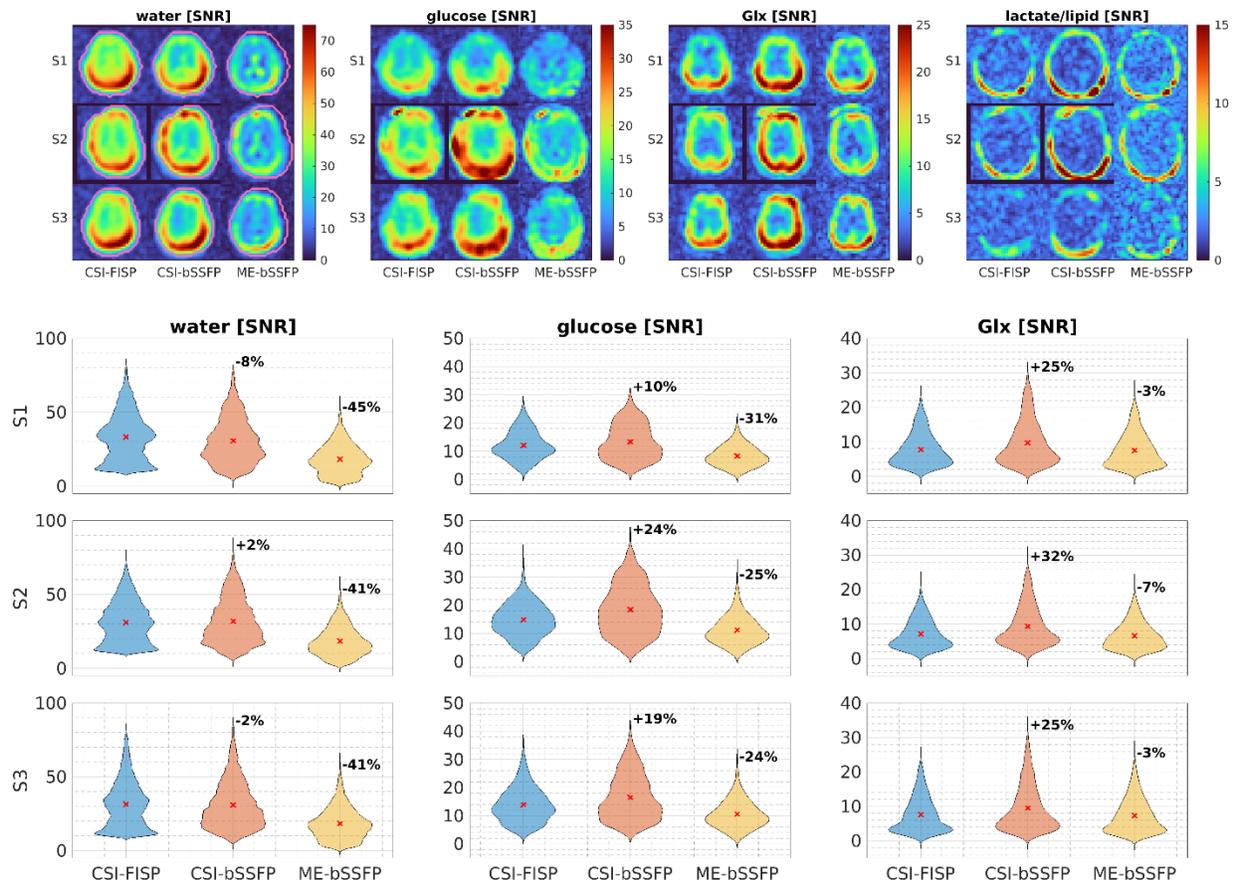

**Figure 5:** *In vivo* SNR analysis for all four deuterated metabolites in all three subjects. An exemplary axial slice of the metabolite maps (top) and the SNR distribution over the entire brain (bottom) are shown for the three different acquisition methods. The whole-brain ROIs are depicted as overlaid magenta contour lines in the water SNR image. The average SNR is illustrated as a red cross and the percentage change of the average SNR relative to the CSI-FISP acquisition is depicted next to the respective violin plots. The interleaved CSI-FISP, CSI-bSSFP and ME-bSSFP maps are averaged to minimize the influence of metabolic dynamics.

In Figure 6, the influence of phase cycling and different methods for combining the acquired phase-cycled data on the SNR is illustrated using the CSI-bSSFP data from subject 1. SNR maps obtained from individual phase cycles (RF phase increments of 180°, 270°, 360° and 90°) and for three different phase cycle combination methods are investigated. The single phase-cycle SNR maps are multiplied by 2 to account for the short acquisition time. The linear and IDEAL-modes methods achieved higher SNR compared with the generic IDEAL algorithm applied to averaged phase cycles. The scaled SNR maps from individual phase cycles show that the measurements with a single phase cycle can achieve higher SNR per unit time at the cost of increased sensitivity to $B_0$ inhomogeneity.

Furthermore, the optimal RF phase increment for different metabolites is not same because of the non-uniform spacing of the chemical shifts.

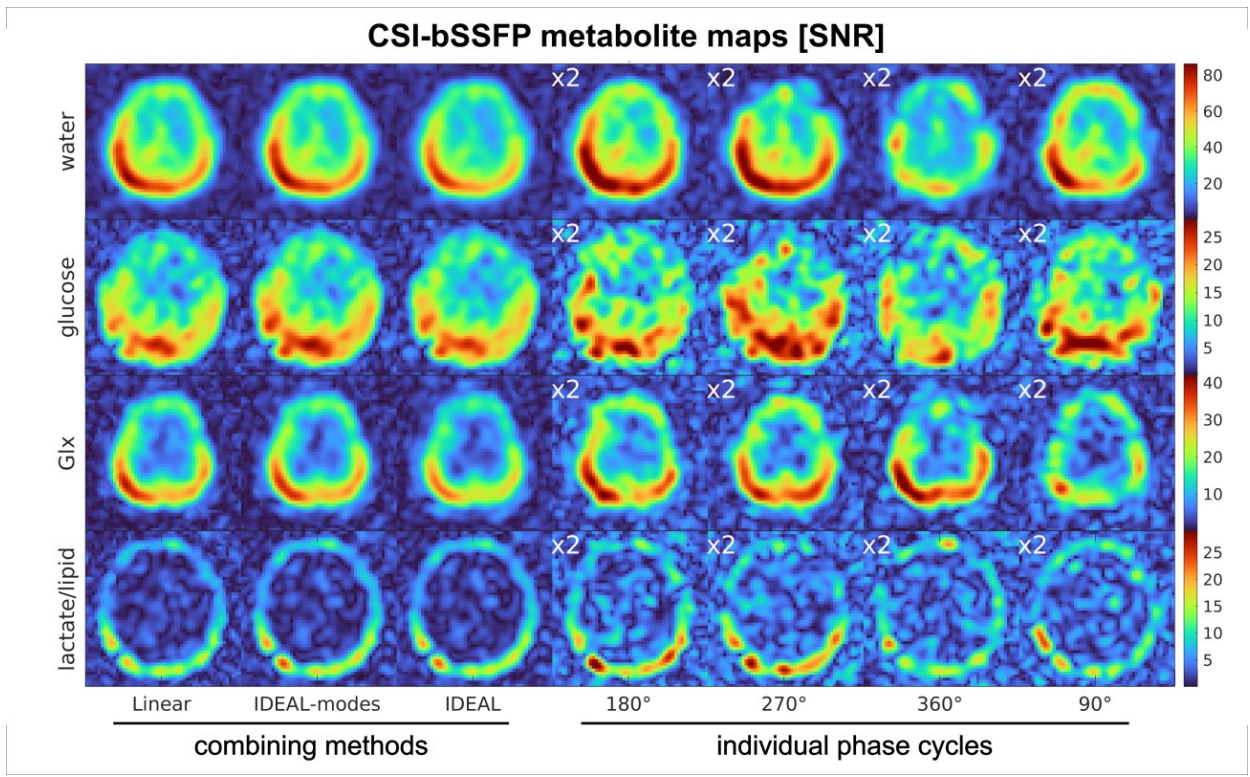

Figure 6: The CSI-bSSFP maps for all metabolites a) water, b) glucose, c) Glx and d) lactate/ lipid. The first three columns show the three different phase cycle combination methods (linear, IDEAL-modes and standard IDEAL algorithm on averaged phase cycles). SNR maps obtained from individual phase cycles (RF phase increments of 180°, 270°, 360° and 90°) are scaled by a factor of two and are shown in the last four columns.

The quantitative results of whole-brain dynamic DMI performed with acquisition-weighted CSI-FISP and CSI-bSSFP acquisitions performed in subject 1 are summarized in Figure 7. The water SNR maps show strong spatial variation due to the receive sensitivity of the surface coils. The quantitative glucose and Glx maps acquired at nominal voxel volume of 0.58 mL show a high degree of spatial localization and minimal coil sensitivity bias. In particular, the expected spatial distribution of Glx in the cortical gray matter regions with higher metabolism is visible. Due to minor acquisition-weighting differences, the actual resolution estimated from PSF is 2.85 mL and 2.98 mL for CSI-FISP and CSI-bSSFP, respectively. Because of phase cycling, there is no visible banding artefacts in the CSI-bSSFP maps. The quantitative metabolite maps of glucose and Glx were obtained using the water image acquired at the earliest time point of the corresponding acquisition technique as reference. Therefore, the water image acquired at 45 minutes and 56 minutes after glucose intake was used as reference for CSI-FISP and CSI-bSSFP maps, respectively. The lower water concentrations in the skull caused artificially inflated glucose concentration estimates in these regions. The temporal evolution of the $^2$H metabolites in the entire brain show excellent agreement between both acquisition techniques evidenced by the alignment of both, the evolution of the average concentration over time and the histogram of metabolite concentrations. The lactate and lipids signals cannot be spectrally separated, which makes it challenging to obtain reliable relaxation times and further accurate quantification of these metabolites.

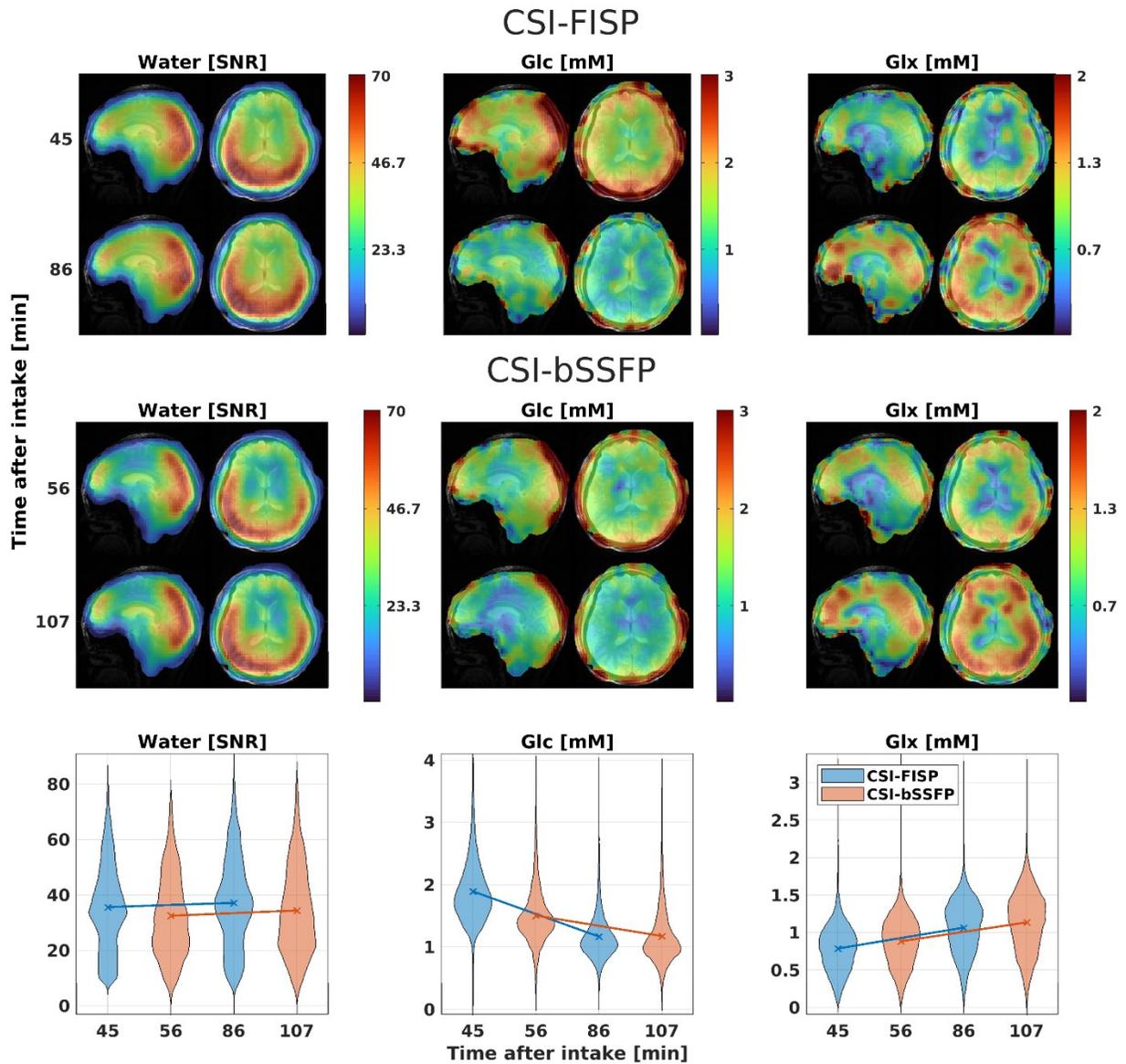

**Figure 7:** Comparison of metabolite maps from the CSI-FISP and CSI-bSSFP acquisitions in subject 1. The 3D maps of three deuterated metabolites (water in SNR units, quantitative glucose and Glx (glutamate+glutamine) maps in mM) measured at different time points after the glucose intake are overlaid over an anatomical reference. The violin plots show the time evolution of $^2$H metabolite across the entire brain.

Similarly, a comprehensive summary of the two acquisition-weighted CSI-FISP and three ME-bSSFP acquisitions is presented in Figure 8. The actual voxel volume of ME-bSSFP of 2.18 mL is much closer to the nominal resolution of 2 mL because of the use of only elliptical scanning with uniform k-space weighting. Therefore, the ME-bSSFP maps exhibit high spatial granularity compared to CSI-FISP. The quantitative glucose maps and the temporal trend of mean concentrations show good agreement between the methods. However, ME-bSSFP Glx maps consistently predict higher concentrations and show a broader distribution of metabolite concentrations compared to both of the CSI counterparts.

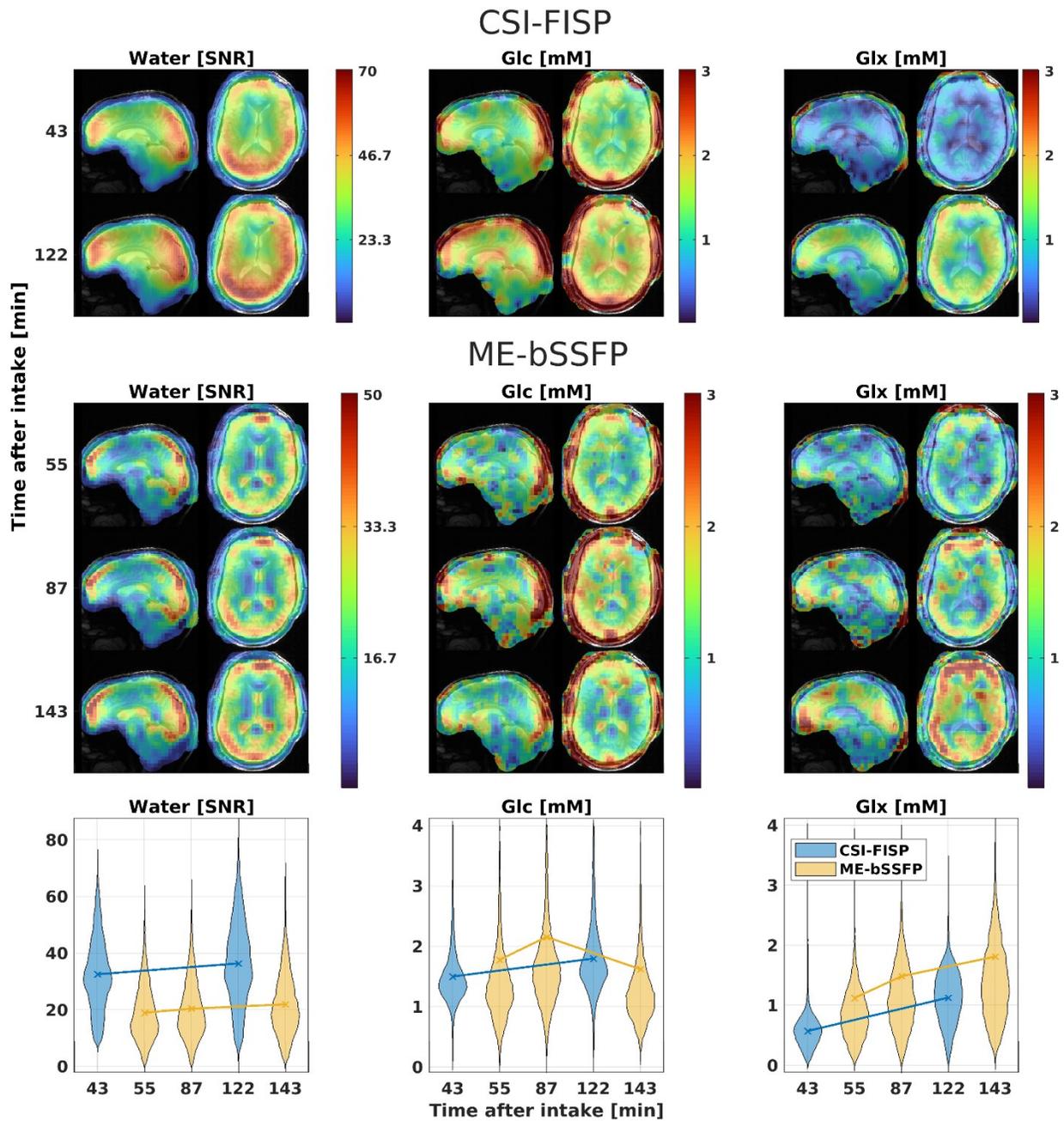

Figure 8: Comparison of metabolite maps from the CSI-FISP and ME-bSSFP acquisitions in subject 2. The 3D maps of three deuterated metabolites (water in SNR units, quantitative glucose and Glx (glutamate+glutamine) maps in mM) measured at different time points after the glucose intake are overlaid over an anatomical reference. The violin plots show the time evolution of $^2$H metabolite across the entire brain.

## Discussion

In this work, we demonstrated the feasibility of using CSI-bSSFP and ME-bSSFP acquisitions with phase cycling for off-resonance insensitive high-resolution [6,6'-$^2$H$_2$]-labeled glucose DMI studies in the healthy human brain at 9.4T. The measured relaxation times of the $^2$H metabolites at 9.4 T were in general agreement with the literature[2,9,10,13]. The ideal Lorentzian line shape assumption in the AMARES fitting of the non-localized spectra resulted in a large uncertainty in the fitted peak amplitudes. This translated to large variations across subjects. The bi-exponential $T_1$ relaxation of *in vivo* water was not apparent in the inversion recovery data. However, the spin-echo data showed strong bi-exponential $T_2$ relaxation and therefore a bi-exponential model was used to fit water peak amplitudes. Utilizing the measured *in vivo* relaxation times, all *in vivo* protocols were optimized to improve the signal level of lower concentration metabolites (glucose and Glx) rather than water (see Figure 2). Both investigated bSSFP variants were set up to acquire multiple phase cycles to achieve reliable metabolite amplitude estimation across the brain. The spectral resolution of all protocols to resolve all $^2$H resonances were verified based on phantom experiments.

Both SNR optimized CSI protocols with acquisition weighting achieved a nominal isotropic resolution 0.58 mL, representing a significant improvement over the nominal resolution of 3 mL achieved in previous studies conducted at the same site[9]. The PSF analysis suggests that the actually resolved volume with acquisition weighting is approximately 5-6 times larger than the nominal voxel volume. Without acquisition weighting, the higher nominal resolution of ME-bSSFP of 2 mL is close to the actual voxel volume of 2.18 mL. Despite SAR restrictions and addition of phase cycling, CSI-bSSFP demonstrated improved average SNR for the detection of glucose (+18%) and Glx (+27%) compared with the vendor's standard CSI-FISP (see Figure 5). The SNR loss of *in vivo* water (-8%) can be attributed to the suboptimal flip angle due to the low $T_2/T_1$ ratio. In phantom experiments, only water and lactate with a large $T_2/T_1$ ratio and large $T_2$ relaxation time showed a clear signal enhancement for CSI-bSSFP in comparison to CSI-FISP (see Figure 4). Notably, the CSI-FISP protocol used in this study already operates in an SNR optimal range and the FISP acquisition achieves higher signal levels than FLASH acquisition with RF spoiling especially for short TRs. Therefore, a 2-3-fold SNR boost for bSSFP relative to gradient spoiled CSI as reported in recent related preclinical work[13] could not be achieved here for CSI-bSSFP vs. CSI-FISP. Even after accounting for the smaller voxel size, ME-bSSFP consistently reported lower average SNR (-43% for water, -27% for glucose, -4% for Glx) compared with CSI-FISP. This discrepancy may be due to the non-uniform bSSFP frequency response near the passband due to the suboptimal flip angle dictated by the SAR limit. The uniformly distributed RF phase increments over 360° likely led to more time spent acquiring phase cycles with lower signal levels, as illustrated in Supporting Information Figure S4.

### Encoding schemes

In general, CSI encoding is optimal for achieving better spectral encoding while the multi-echo encoding is optimal for achieving higher spatial resolution. Because of the poor sensitivity of DMI and sparse $^2$H spectrum, both encoding schemes fulfil the low spectral-spatial resolution requirements of DMI. CSI is typically the preferred method because it offers a higher ADC duty cycle due to less gradient switching, silent operation and better interpretable spatially resolved frequency spectra. The fast acquisition of ME-bSSFP could be advantageous in case signal averaging is used for measuring more phase cycles as investigated in this work.

### Phase cycling

The combination of a relatively long TR of 19 ms, required to achieve sufficient spectral dispersion, and significant $B_0$ inhomogeneity in the human head at 9.4 T (3.5 ppm≈210 Hz) necessitated the use of phase cycling to minimize off-resonance sensitivity. The employed relatively high number of eighteen phase cycles in ME-bSSFP, mitigated aliasing in higher order modes and provided additional spectral encoding. As some averages are required for acquisition weighting, only four phase cycles were used in the CSI-bSSFP protocol. Since the amplitudes of the higher order SSFP modes decay fast due to short $T_2$ relative to TR and large flip angles[19], no visible aliasing of higher order modes was observed *in vivo* even with only four phase cycles.

In addition to the chemical-shift induced phase evolution, the frequency-dependent amplitude and phase modulation along phase cycles provides complementary spectral information. Although all protocols in this study are designed to resolve all four metabolites without phase cycling, the additional spectral encoding from the phase cycling could be used to reduce the TR. The spectral conditioning can be further improved with phase cycling by suppressing specific resonances by adjusting TR to shift them to the bSSFP frequency response stopband.

The disadvantage of using phase cycling is essentially the SNR loss due to the low signal in the bSSFP stopband, which is substantially lower than the FISP steady-state signal for the assessed *in vivo* relaxation times and employed sequence parameters (see Supporting Information Figure S4). Otherwise, the time penalty of phase cycling is negligible since signal averaging already requires multiple repetitions, and the transition period between phase cycles is short.

## Spectral Separation

In addition to improving the signal levels through bSSFP acquisition, it is essential to minimize noise amplification during the spectral separation to retain the SNR advantage. This can be achieved by selecting an appropriate signal model that accurately explains the acquired signal and by providing proper conditioning during the signal model inversion. A signal model that incorporates phase cycling, relaxation times and $B_0$ off-resonance in addition to the chemical shifts can describe the acquired signal accurately. However, to ensure proper conditioning, the echo spacing of the multi-echo protocols and readout lengths of the CSI protocols are chosen to maximize the number of signal averages[20] (Fisher information) for all $^2$H resonances (see Supporting Information Figure S5).

The IDEAL algorithm used for CSI-FISP data fits the signal model by considering only chemical shifts and $B_0$ off-resonance. In case of phase-cycled bSSFP data, the more descriptive model of the data used in the proposed linear fitting method achieves higher SNR than simply averaging the phase cycles (see Figure 6). As this model includes the amplitude and phase variations due to phase cycling, it has the added benefit of providing additional spectral encoding information as shown in Figure 3. However, due to the uncertainty in the estimation of the relaxation times and flip angle, the IDEAL-modes algorithm employing a simplified signal model, which considers only the phase evolution of the metabolites across SSFP configurations is proposed in this work. It further improves SNR in comparison to the linear fitting method by reducing any bias in the signal model through implicit determination of the relaxation times and flip angles from the data. The optimal weights calculated with Eigen decomposition for combining metabolite amplitudes across SSFP modes show close resemblance to the simulated SSFP mode decay of individual metabolites (see Supporting Information Figure S6). As only the chemical shift and $B_0$ off-resonances are fitted to individual modes, the spectral information in the phase cycles are not utilized in this method. On the other hand, the flexibility of the simple model makes it suitable for application to other field strengths or different tissue types (e.g. tumors) with unknown relaxation times or chemical species with relaxation time heterogeneity such as water in brain tissue and ventricles.

## Metabolite Quantification

The obtained quantitative metabolite maps were of high quality, without any major artifacts, indicating sufficient SNR throughout the entire brain, even at relatively small voxel sizes. To maximize the data quality for reliable SNR comparison, the acquisitions were performed 30 minutes after glucose intake. Therefore, the assumed 10.12 mM water reference is in fact slightly higher and results in a minor underestimation of glucose and Glx concentrations. The strong agreement of metabolite concentrations between different methods suggests accurate calibration factors calculated using the respective signal model assumptions of the different acquisition techniques. The ME-bSSFP metabolite map demonstrated higher dispersion of metabolite concentrations, especially for Glx, indicating higher resolution and reduced partial volume dilution effects compared with the acquisition-weighted CSI acquisitions.

All sequences achieved good SNR for glucose allowing the detection of increased glucose uptake in the skull in addition to the brain as shown in a recent high-resolution FDG-PET study[35]. The raw glucose maps from the ME-bSSFP acquisition showed highly localized uptake in the skull (see Supporting Information Figure S7). In some subjects, the stronger glucose signal in the skull than in the brain may be due to a combination of late measurement time, receiver sensitivity bias, and rapid metabolic conversion of glucose in the brain.

## System consideration

The reference power was set lower due to the limited RF peak power of approximately 4 kW (447 V peak amplitude) and the vendor's convention of defining reference power with a 0.5 ms RF pulse duration. As a result, the actual flip angles achieved were 85-90% of the nominal flip angles. The limited power availability also led to longer inversion and refocusing pulses in the relaxometry experiments. A small variation in $B_1^+$ efficiency of 147 - 160 nT/V, calculated from the reference amplitude measurements, was observed across different head sizes. Given the simulated maximum local SAR ($SAR_{10g}$) of 0.296 W/kg for 1 W input power, safety margins due to simulation inaccuracies, and increased measurement uncertainty of directional couplers around 61 MHz, the allowed maximum instantaneous time-averaged power was approximately 12 W for all sequences lasting more than 6 minutes. Therefore, the pulse duration of the bSSFP acquisitions was increased to 1.4 ms to achieve high flip angles. On the receiver side, adding a second-stage amplification of 18 dB (high gain option) to the 26 dB power gain of the pre-amplifier significantly improved SNR, indicating a dominant noise in the receive chain than the MR signal.

## Conclusion

This work successfully developed two bSSFP acquisition and processing methods for performing DMI at a human 9.4 T scanner for investigating healthy human brain. All experiments were performed at an improved spatial resolution of 2-3 mL (equivalent of acquisition-weighted 0.4-0.6 mL nominal resolution) and high temporal resolution of 10 minutes. Balanced SSFP acquisition has potential to improve the sensitivity of DMI despite the SNR loss of phase-cycling and other human scanner constraints.

## Acknowledgments

The authors thank Dario Bosch for the insights into MR safety and online SAR monitoring. They also thank Felix Glang for his valuable advice on SNR analysis. Financial support from the Max-Planck-Society, Deutsche Forschungsgemeinschaft (the Reinhart Koselleck Project) (SCHE658/S2), the European Research Council (Spread MRI) (Grant 834940), and Bundesministerium für Bildung und Forschung (01GQ1805B) is gratefully acknowledged. We also acknowledge support from the Deutsche Forschungsgemeinschaft (DFG, German Research Foundation) by the Cluster of Excellence iFIT (EXC 2180) "Image-Guided and Functionally Instructed Tumor Therapies," University of Tuebingen, Germany (KS and AFM), the Werner Siemens Foundation, the Alexander von Humboldt Foundation within the framework of the Sofja Kovalevskaja Award (AFM), and the DKTK German Cancer Consortium Innovation program "HYPERBOLIC" (LK).

## DATA AVAILABILITY STATEMENT

The MATLAB software package used for simulation and DMI processing are available on Github at https://github.com/praveenivp/DeuteMetcon.git (v0.1.0). The Pulseq sequences and associated processing scripts for estimating $T_1$, $T_2$ and reference voltage were also shared in the same repository. The DMI raw data from both phantom and subjects as well as non-localized spectroscopy data is openly available in zenodo: https://zenodo.org/doi/10.5281/zenodo.14652737 . The instruction manual for phantom preparation and the measurement protocols are also shared along with the data repository. Sequence binaries can be obtained through the Siemens Healthineers customer-to-customer partnership program (C2P procedure) upon reasonable request.

# Supporting Information

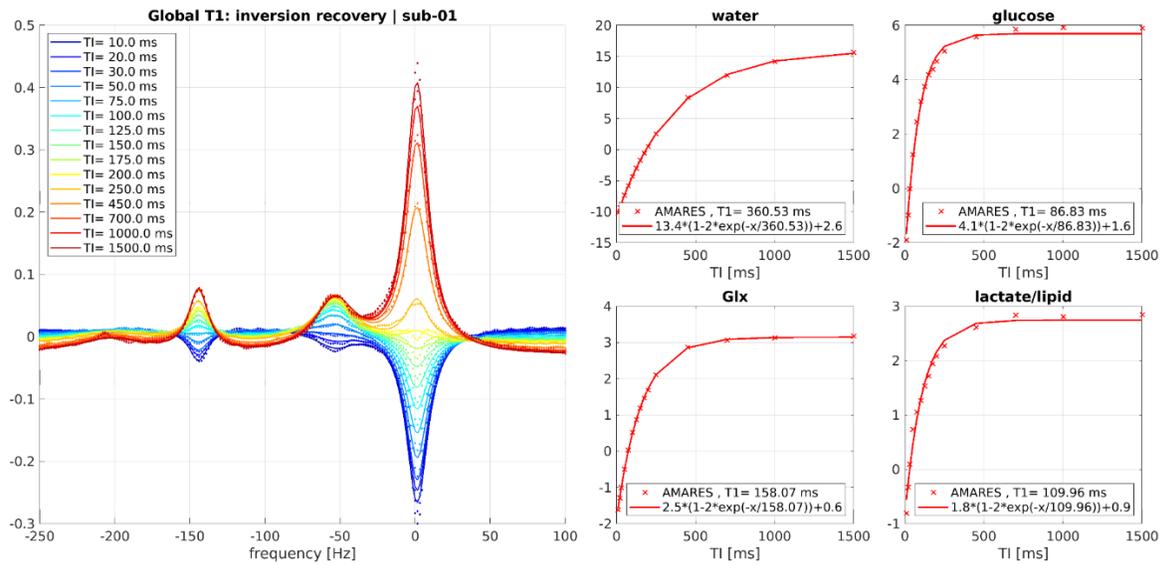

**Supporting Information Figure S1:** Non-localized inversion recovery $T_1$ measurements performed approximately 90 minutes after glucose intake in subject 1. The AMARES fittings performed for all the inversion times are shown on the left along with the data. The exponential fits performed on the amplitudes of the individual $^2H$ resonances are shown on the right.

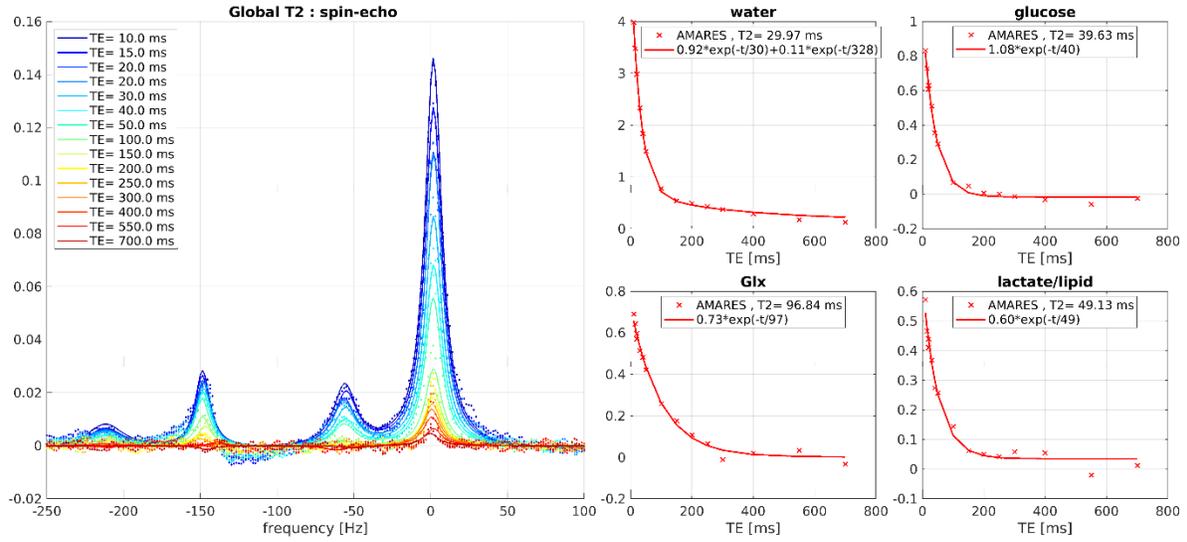

**Supporting Information Figure S2:** Non-localized spin-echo T$_2$ measurements performed approximately 100 minutes after glucose intake in subject 1. The AMARES fittings performed for all the echo times are show on the left along with the data. The exponential fits performed on the amplitudes of the individual $^2$H resonances are shown on the right.

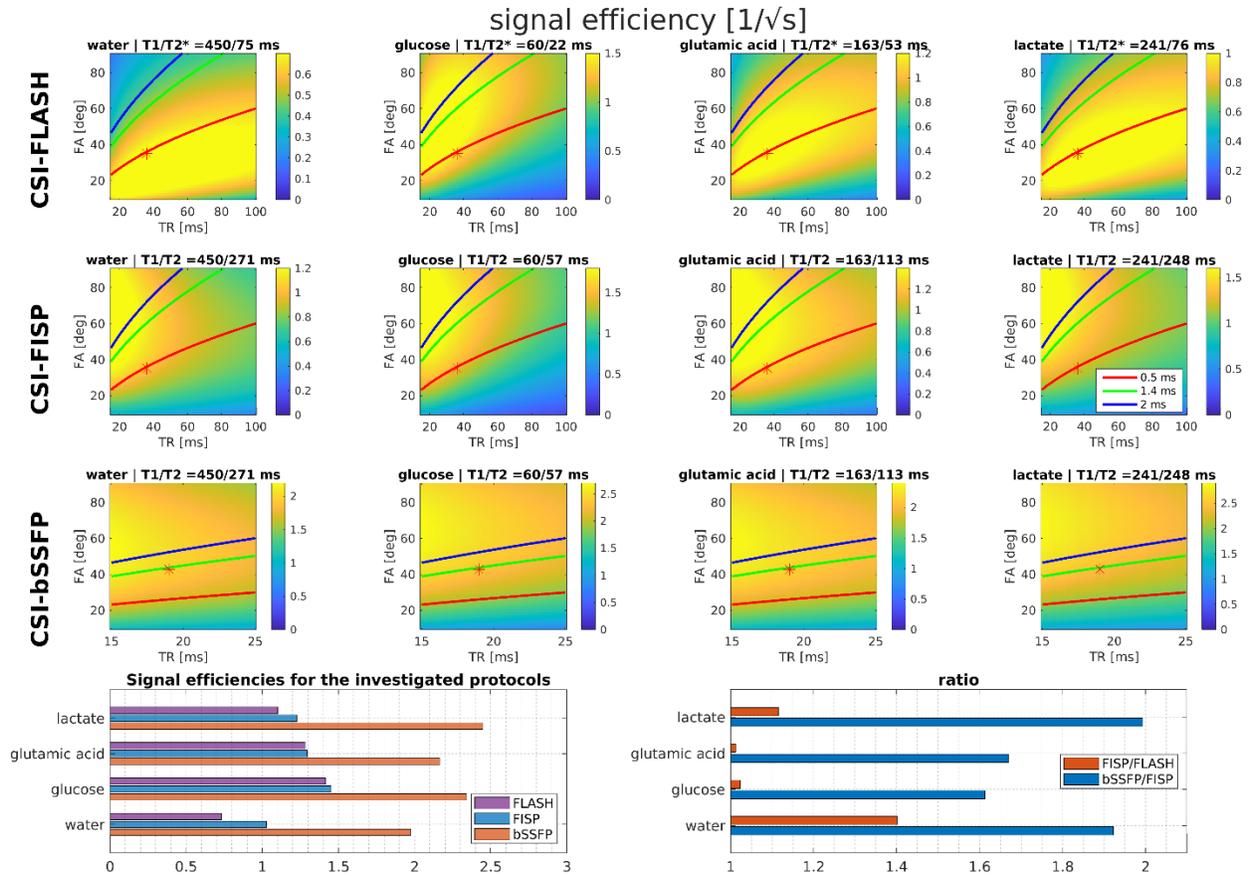

**Supporting Information Figure S3:** In vitro SNR efficiency (signal amplitude/$\sqrt{TR}$) of deuterium metabolites with respect to TR and flip angle for FLASH, FISP and bSSFP acquisitions. The measured in vitro relaxation times used for the simulation are shown in the title of the respective subplots. The SAR limits with respect to TR for three pulse durations are overlaid to show the available parameter space for SNR optimization. The asterisks in the plots indicate the parameter combination (flip angle, TR) of the protocols used in this study. The signal efficiency and their ratios of the three protocols for all four metabolites are depicted in the bottom panels.

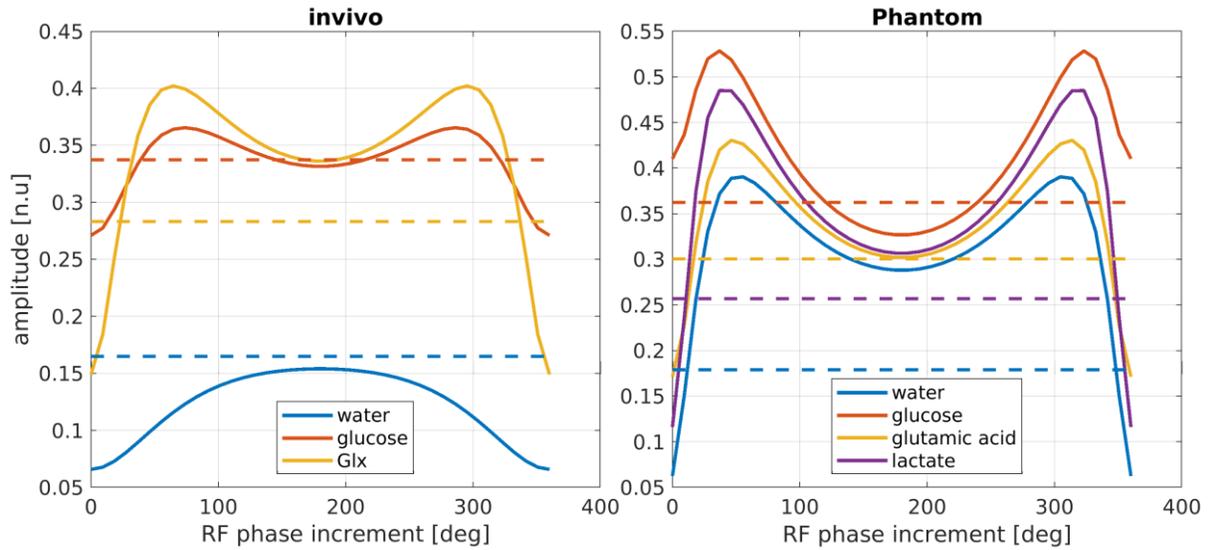

**Supporting Information Figure S4:** The in vivo bSSFP signal level (solid line) of three metabolites for different RF phase increments is shown on the left along with the FISP signal level (dotted line). The signal levels are simulated for the protocols used in this study and the measured in vivo relaxation times at 9.4 T. Similarly, the bSSFP signal level along with the FISP signal level is shown on the right for all four metabolites with the relaxation times measured in the phantom.

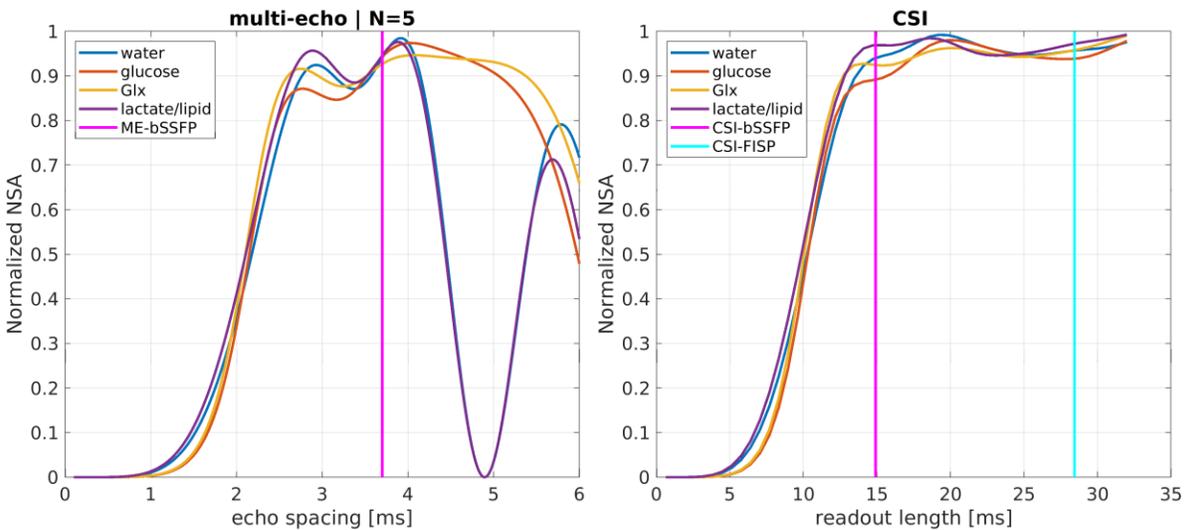

**Supporting Information Figure S5:** The normalized number of signal averages (NSA) or the Fisher information estimated from the system model matrix with four $^2$H resonances at different echo spacings for the multi-echo acquisition and different readout lengths for the CSI acquisitions. The magenta and cyan line markers show the echo-spacing and readout lengths used in the optimized protocols of this study.

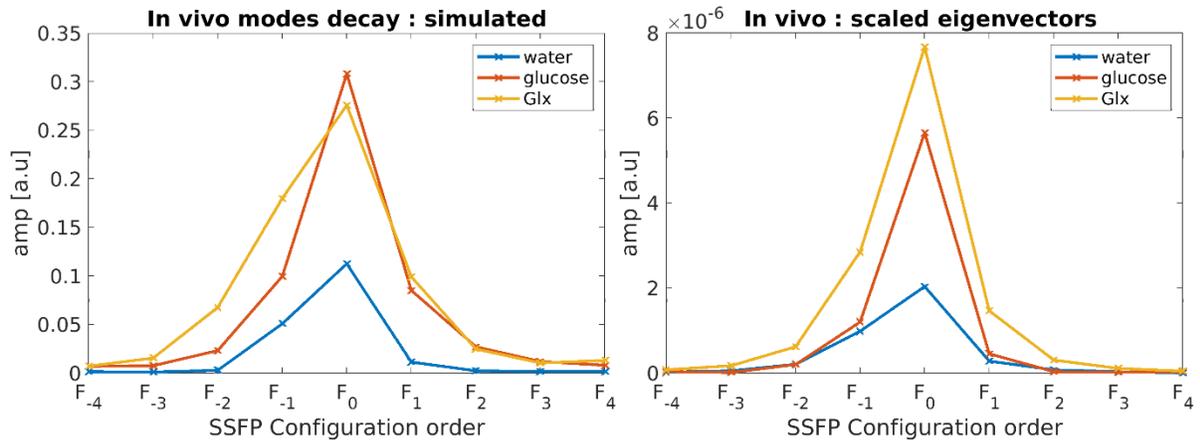

**Supporting Information Figure S6:** The SSFP mode amplitudes estimated from the simulated bSSFP frequency response for the measured in vivo relaxation times and a 50 deg flip angle. The principal eigenvectors of all three metabolites estimated from the fitted metabolite amplitudes across different modes.

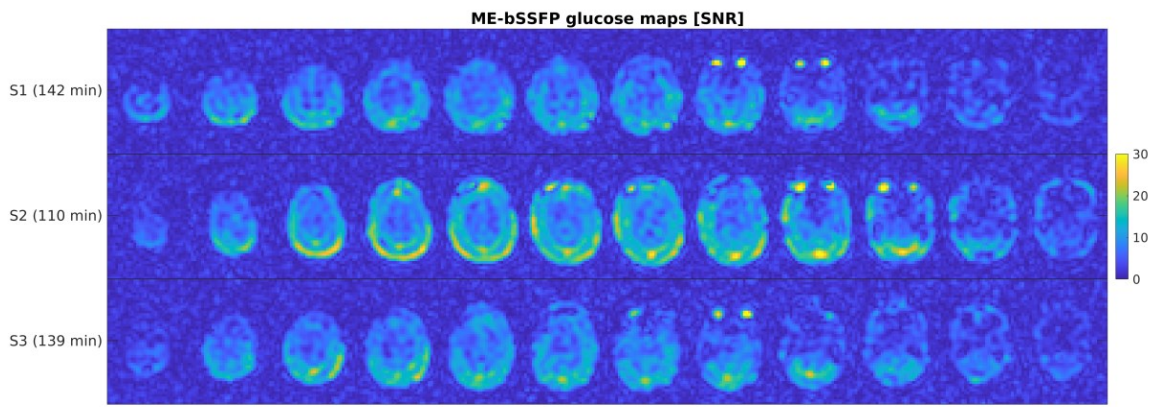

**Supporting Information Figure S7:** The axial slices of ME-bSSFP glucose maps for all subjects showing highly-resolved pronounced uptake of glucose in the skull. The acquisition time after the glucose intake is shown for each subject. The maps are not corrected for receive sensitivities.